\documentclass[twocolumn,aps,showpacs,superscriptaddress]{revtex4}

\usepackage{amssymb}
\usepackage{amsfonts}
\usepackage{amsmath}
\usepackage{graphicx}
\usepackage{bm}
\usepackage{multirow}

\newcommand{\vect}{\boldsymbol}
\newcommand{\colvect}[2] {\begin{pmatrix} #1 \\ #2\end{pmatrix}}
 
\newcommand{\smfrac}[2]{\tfrac{ #1 }{#2}}

\begin{document}

\title{Dirac edges of fractal magnetic minibands in graphene with hexagonal moir\'e superlattices}

\author{Xi Chen}
\affiliation{Department of Physics, Lancaster University, Lancaster, LA1 4YB, UK}
\author{J.~R.~Wallbank}
\affiliation{Department of Physics, Lancaster University, Lancaster, LA1 4YB, UK}
\author{A.~A.~Patel}
\affiliation{Department of Physics, Harvard University, Cambridge MA 02138, USA}
\author{ M.~Mucha-Kruczy\'nski}
\affiliation{Department of Physics, University of Bath, Claverton Down, Bath, BA2 7AY, UK}
\author{E.~McCann}
\affiliation{Department of Physics, Lancaster University, Lancaster, LA1 4YB, UK}
\author{V.~I.~Fal'ko}
\affiliation{Department of Physics, Lancaster University, Lancaster, LA1 4YB, UK}

\date{\today}

\begin{abstract}
We find a systematic reappearance of massive Dirac features at the edges of consecutive minibands formed at magnetic fields $B_{p/q}\!=\!\smfrac{p}{q}\phi_0/S$ providing rational magnetic flux through a unit cell of the moir\'e superlattice created by a hexagonal substrate for electrons in graphene. The Dirac-type features in the minibands at $B\!=\!B_{p/q}$ determine a hierarchy of gaps in the surrounding fractal spectrum, and show that these minibands have topological insulator properties. 
Using the additional $q$-fold degeneracy of magnetic minibands at $B_{p/q}$, we trace the hierarchy of the gaps to their manifestation in the form of incompressible states upon variation of the carrier density and magnetic field.
\end{abstract}


\maketitle 
The fractal spectrum of electron waves  in crystals subjected to a strong magnetic field \cite{zak_physrev_1964,brown_physrev_1964} is a fundamental result in the quantum theory of solids \cite{stat_phys_LL}. 
In particular, the bandstructure of electrons on a two-dimensional (2D) lattice is fractured into multiple bands which occur at the values $B_{\frac{p}{q}}=\frac{p}{q}\phi_0/S$ of the field providing a rational fraction of magnetic flux quantum $\phi_0=h/e$ per unit cell area $S$ \cite{zak_physrev_1964,stat_phys_LL}.
Its image \cite{hofstadter_prb_1976}, obtained for a square lattice hopping model, known as the Hofstadter butterfly, has stimulated numerous attempts to observe the fractal spectrum of electrons via quantum transport measurements.
Since the sparsity of the spectrum increases for larger values of the denominator $q$ in $\smfrac{p}{q}$ (hence smaller gaps), the observation of fractal magnetic bands in real crystals would require unsustainably strong magnetic fields. 
Hence, early efforts were focused on 2D electrons in periodically 
patterned GaAs/AlGaAS heterostructures \cite{semiconductor_superlattices}, where the superimposed superlattice period was made
large enough to obtain low denominator fractions within the experimentally available steady magnetic field
range. The more recent observation of moir\'e superlattices, both for twisted bilayer graphene \cite{twisted_blg} and graphene residing on substrates 
with hexagonal facets \cite{moire_metals}, has shown an alternative way to create a long-range periodic potential 
for electrons: by making lattice-aligned graphene heterostructures using a hexagonal crystal with an almost commensurate period. 
For this, hexagonal boron nitride (hBN, with the lattice constant of $l_{hBN}=2.50\,\text{\AA}$ vs $l=2.46\,\text{\AA}$ in graphene) provides a perfect match. Several observations of
moir\'e superlattice effects in graphene-hBN 
heterostructures have already been reported \cite{ponomarenko_nature_2013,dean_nature_2013,hunt_science_2013}.

In this article, we study magnetic minibands caused by a moir\'e superlattice in graphene. When the surface layer of the substrate is inversion symmetric, the zero-magnetic-field spectra displays clear secondary Dirac points at the first miniband edge \cite{wallbank_prb_2013, ortix_prb_2012, kindermann_prb_2012}. 
We find that generations of massive Dirac electrons systematically reappear at the edges of magnetic minibands at $B \!=\! B_{\frac{p}{q}}$, and the surrounding fractal spectrum groups around $q$-fold degenerate Landau levels (LLs) of gapped Dirac electrons in an effective magnetic field $\delta \! B \!=\! B - B_{\frac{p}{q}}$. 
As the Dirac model features a ``zero-energy'' LL, separated by the largest gap from the rest of 
the spectrum, this determines a specific hierarchy of gaps in the Hofstadter butterfly, resulting in 
a peculiar sequence of dominant incompressible states of electrons in graphene-hBN heterostructures in a strong magnetic field.

	\begin{figure}[tb]
	\centering
	\includegraphics[width=0.48 \textwidth]{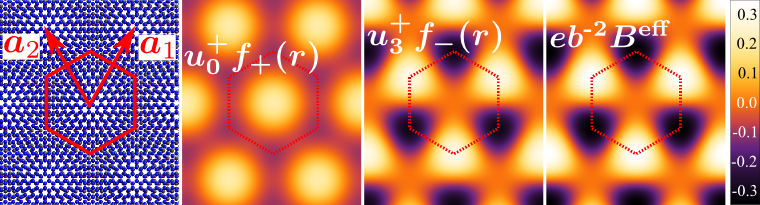}
	\caption{
	Moir\'e pattern ($\vect a_1\!=\!a(\smfrac{1}{2},\smfrac{\sqrt{3}}{2})$, $\vect a_2\!=\!a(\smfrac{-1}{2},\smfrac{\sqrt{3}}{2})$, unit cell area $S\!=\!\smfrac{\sqrt3}{2}a^2$) and effective perturbations for electrons in graphene on an hexagonal substrate [$\theta\!=\!0$, and  $V^+\!=\!0.063vb$, $V^-\!=\!0$ in Eqs.~(\ref{eq:H}-\ref{eq:model})].}
	\label{fig:moire}
	\end {figure}
 
\begin{figure}[tb]
	\centering
	\includegraphics[width=0.45\textwidth]{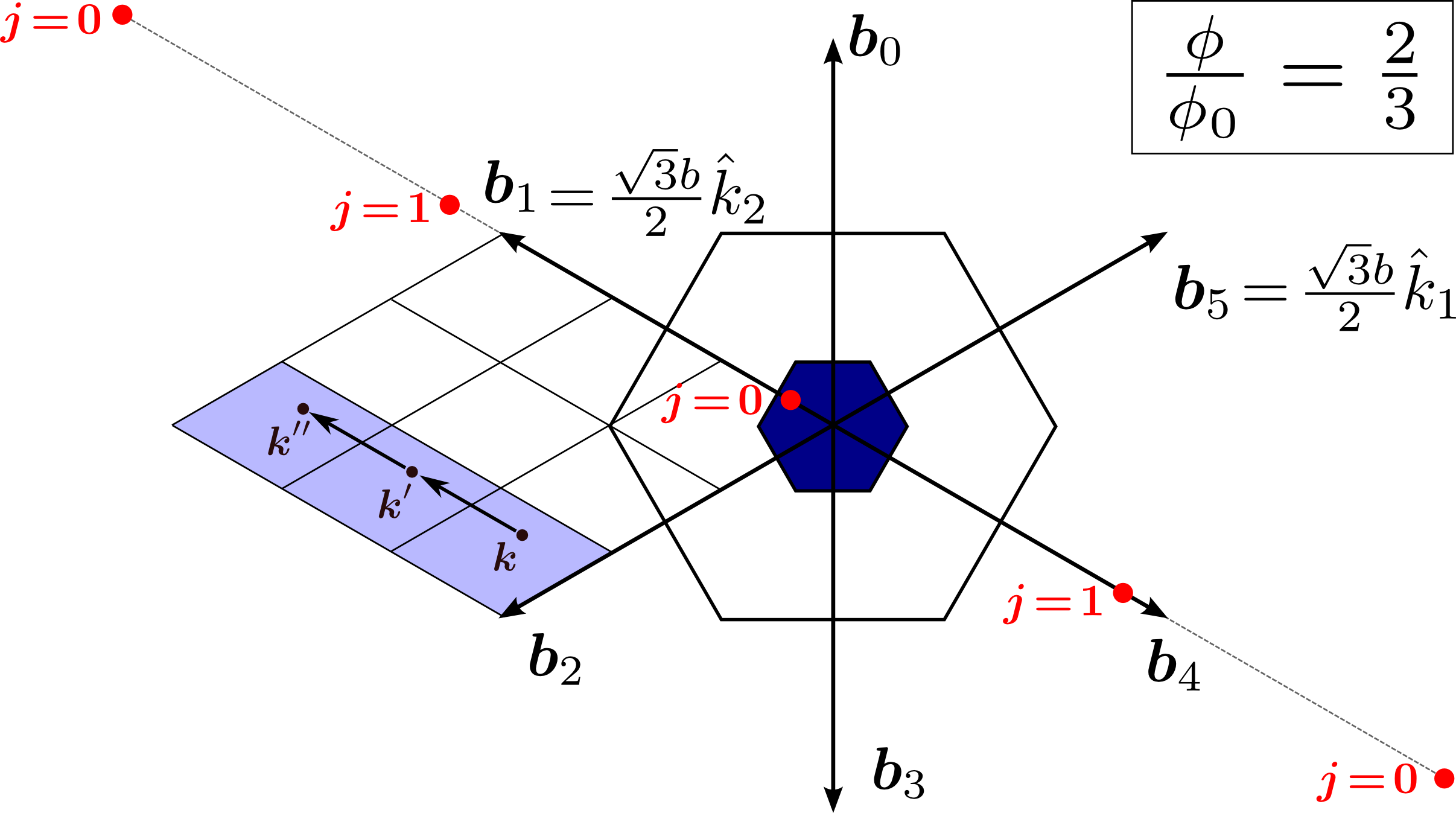}
	\caption{
	Example of the moir\'e superlattice Brillouin zone (larger hexagon, with area $\sqrt{3}b^2/2$, and reciprocal lattice vectors $\vect b_m$), and the two smaller magnetic Brillouin zones (shaded blue, discussed in the text). Red dots show $k_2$ values used to construct Bloch states $|_{ t, \vect k}^{ n,j}\rangle$.  	
}
	\label{fig:magnetic_BZ}
	\end {figure}

For graphene electrons we use the effective Dirac theory, where  the form of moir\'e perturbation depends on the detailed structure and symmetry  of the underlying surface. 
The phenomenological model for the moir\'{e} superlattice \cite{wallbank_prb_2013} is given by the Hamiltonian, 
   \begin{align}
  \!\hat H=&v \vect p \!\cdot \!\vect\sigma+
 v b\!\left( u_0^+  f_+ \! +\! u_0^-   f_- \right) 
 + \xi v b \sigma_3\!\left(u_3^+   f_-   \! +\!u_3^-   f_+  \right)\nonumber\\
 &+ \xi v \vect \sigma \!\cdot\! \left[\vect l_z\!\times\!\nabla\! \left( u_1^+  f_-  \!+ \! u_1^-   f_+ \right) \right]\!. \label{eq:H}
 \end{align}
Here $\sigma_i$ are Pauli matrices, acting on Bloch states $(\phi_{AK},\phi_{BK})^T$ in the $K$ valley ($\xi=1$) and $(\phi_{BK'},-\phi_{AK'})^T$ in the $K'$ valley ($\xi=-1$).
In the Dirac term, $\vect p\!=\!-i\nabla+e \vect A$, ($\hbar=1$) describes
momentum relative to the valley center, and $[ \nabla \times\vect A]_z=B$. 
The rest of Eq.~\eqref{eq:H} describes the superlattice perturbation: a simple potential modulation,  a local $A$-$B$ sublattice asymmetry, and the variation of the $A$-$B$ hopping mimicking a pseudo-magnetic field $e B^{\text{eff}}=\xi b^2 (u_1^+ f_-  + u_1^- f_+)$. 
Here, $f_\pm =\!\sum_{m}\!(\pm1)^{m+\frac{1}{2}}e^{i\vect b_m\cdot\vect r}$, where the reciprocal lattice vectors, $\vect b_{m=0,\cdots,5}$, responsible for the dominant contribution to the superlattice perturbation \cite{twisted_blg,ortix_prb_2012,kindermann_prb_2012,wallbank_prb_2013}, are related by $60^\circ$ rotations (Fig.~\ref{fig:magnetic_BZ}) with $b\equiv|\vect b_m|\approx \frac{4\pi}{\sqrt{3} l}\sqrt{\delta^2+\theta^2}$, ($\delta=l_{hBN}/l-1\approx 2\%$  and $\theta\ll1$ is the misalignment angle).  
 
Six dimensionless parameters  $u_i^\pm$ in Eq.~\eqref{eq:H}   describe the 2D inversion symmetric/asymmetric part of the moir\'e perturbation.
They can take arbitrary values, but two microscopic models, one based on the hopping between the graphene and hBN lattices \cite{kindermann_prb_2012}, and the other on scattering of graphene electrons off quadrapole electric moments  in the  hBN layer \cite{wallbank_prb_2013}, predict that
\begin{align}
&vb\{u_{i=0,1,3}^\pm\}= V^\pm \left\{\frac{\pm1}{2} , \frac{- \delta}{\sqrt{\delta^2\!+\!\theta^2}},\frac{-\sqrt 3}{2}\right\}.\label{eq:model} 
\end{align}
We also assume that $|V^+|>|V^-|$, since only one of the two atoms, B or N, would produce the dominant effect, thus making the moir\'e perturbation almost inversion-symmetric.
In the absence of a magnetic field, the inversion-symmetric perturbation prescribes a gapless miniband spectrum, with a secondary Dirac-point (DP) either at the edge of the first miniband, or embedded into a continuous spectrum at higher energies, whereas the asymmetric part may open gaps both at the zero-energy DP, $\Delta_0$, and  the secondary DP, $\Delta_1$, on the conduction/valence band side ($s=\pm1$),
\begin{align}\label{eqn:gap_mlg}
&\Delta_0=24vb|u^+_{1}u^-_{0}+u^+_{0}u^-_{1}|,\nonumber\\
&\Delta_1=\sqrt3 |  u^-_0+2s\xi u^-_1-\sqrt 3 \zeta u^-_3|.
\end{align}

To adapt the analysis of the electron spectrum in a magnetic field to the hexagonal symmetry of the moir\'e pattern, we shall use a non-orthogonal coordinate system,
$ \vect r=x_1 \frac{\vect a_1}{a} +x_2 \frac{\vect a_2}{a}$, (see Fig.~\ref{fig:moire}), and the Landau gauge, $ \vect A\!=\!B x_1(-\vect a_1 +2\vect a_2)/(\sqrt3 a)$. Then, the ``free'' massive Dirac electrons and the basis of LL states are described by
\begin{align}\label{eq:H_LL}
&H_D= \left(
\begin{array}{cc}
  \smfrac{1}{2}\xi\Delta  & v d^{\dagger} \\
v d &  \smfrac{-1}{2} \xi\Delta
\end{array}
\right),\\
&d=\frac{-2}{\sqrt{3}}\left[ \partial_{x_1}e^{ i\frac{ 2 \pi}{3}} +  (\partial_{x_2}+  i\sqrt{3} e B x_1/2)e^{ -i\frac{ 2 \pi}{3}}\right]\!,\nonumber\\
&\psi_{0}^{k_2}\!=\! \frac{e^{ik_2 x_2}}{ \sqrt L}\!\colvect{\frac{1-\beta}{2}\varphi_0}{\frac{1+\beta}{2}\varphi _0}\! ,\quad
\psi_{n\neq0}^{k_2}\!=\!  \frac{e^{ik_2 x_2}}{\sqrt{2L} }\!\colvect{\varphi _{n^{\!-} }}{c_n\varphi _{ n^{\!+}  }}\!, \nonumber\\
&\varphi_n \!=\!     A_n   e^{-\frac{z^2}{2}+\beta \frac{i z^2}{2 \sqrt{3}}}\mathbb{H}_n(\!z\!),\quad z\!=\frac{\sqrt{3} x_1}{2\lambda_B}+\beta k_2\lambda_B. \nonumber
\end{align}
 Here, $\beta\!=\!B/|B|$, $A_n\!=\!\sqrt{ 3  }/(\sqrt{n!2^{(n+1)} \lambda_B\sqrt{\pi}} )$, $n^\pm\!=\!|n|\pm(\beta\mp1)/2$,
$c_n\!=\! \beta e^{i\frac{2\pi}{3}} \lambda _B (E_n-\xi  \Delta/2  )/(v\sqrt{ 2|n|})$, $\mathbb{H}_n$ is the Hermite polynomial, and $\lambda_B\!=\!1/\sqrt{|eB|}$  is the  magnetic length. The LL spectrum of the gapped Dirac model is
\begin{align}\label{eq:gapped_En}
&E_{0}=  \frac{-\beta \xi \Delta}{2},\qquad  
E_{n\neq0}=  \text{sign}(n)\sqrt{  \frac{\Delta ^2}{4}  + \frac{ 2|n|v^2}{\lambda^2_B}}.   
\end{align}

For a magnetic field, $B_{\frac{p}{q}}\!=\!\frac{p}{q}\phi_0/S$, the electronic spectrum can be described in terms of minibands of Wannier states on a $q$-times larger period superlattice with an integer number of flux quanta per super-cell. 
For the unit cell enlarged $q$-times \cite{zak_physrev_1964,stat_phys_LL}, e.g., in the $\vect a_1$ direction, the Brillouin Zone (BZ) is $q$-times smaller than the moir\'e BZ (Fig.~\ref{fig:magnetic_BZ}), with
the dispersion repeated inside it with the period  $\vect b_1/q$. 
To preserve the point group symmetries, here we enlarge the unit cell $q$-times in all directions, folding the momentum space on to a magnetic BZ with area $\sqrt{3} b^2/(2q^2)$, where each energy band is $q$-fold degenerate \cite{brown_physrev_1964}. 

\begin{figure*}[t]
\center
\includegraphics[width=0.98\textwidth]{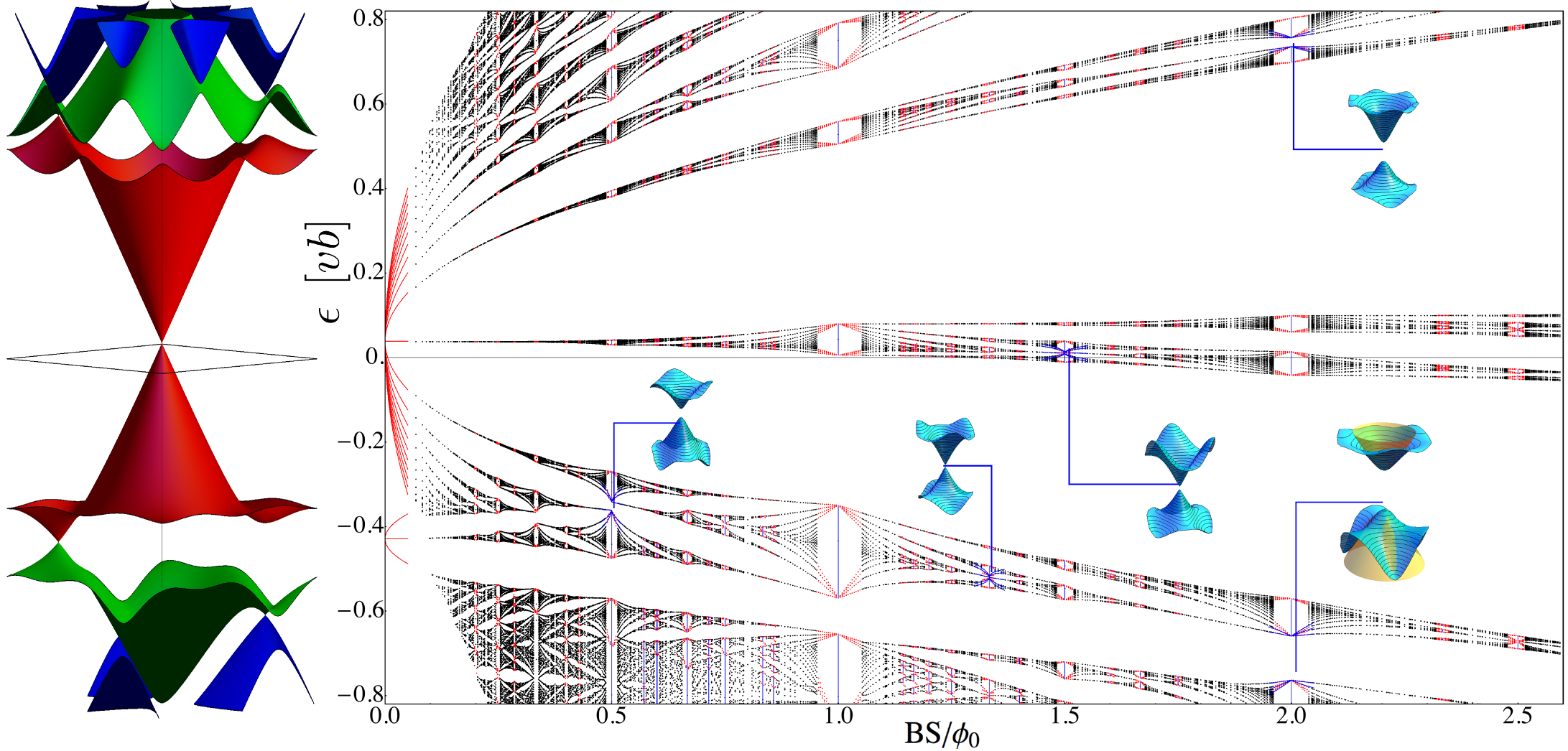} 
\caption{
Fractal energy spectrum found by exact numerical diagonalization. The insets show examples of magnetic minibands, at $B_{p/q}\!=\!\smfrac{p}{q}\phi_0/S$, and their fit (in yellow) with Dirac spectra used to calculate the energies of effective Dirac Landau levels. The panel on the left shows the $B\!=\!0$ minibands with a distinct secondary Dirac point in the valence band.
}
\label{fig:miniband_spectra}
\end{figure*}

The degeneracies of magnetic minibands can be deduced from the properties of the group, $G_M=\{\Theta_{\vect X}\equiv e^{ieBm_1a\frac{\sqrt3}{2}x_2} T_{\vect X}, \vect X=m_1\vect a_1+m_2\vect a_2\}$, of magnetic translations, where  $T_{\vect X}$ are the geometrical translations, and
\begin{gather*}
\Theta_{\vect X}\Theta_{\vect X'} =   e^{i 2\pi\frac{p}{q} m_1'm_2} \Theta_{\vect X+\vect X'}, \nonumber\\
\Theta_{\vect X}\Theta_{\vect X'} =      e^{i2\pi\frac{p}{q}   (m_1'm_2-m_1m_2') }  \Theta_{\vect X'}\Theta_{\vect X}.\label{eq:mag_tran}
\end{gather*}

The subgroup of $G_M$ made of translations $ \vect R=m_1q\vect a_1+m_2q\vect a_2 $ on a $(q\times q)$-enlarged superlattice is isomorphic to the simple group of translations, $T_{\vect R}$, so that its eigenstates, $ \Theta_{\vect R}|_{t, \vect k}^{n,j}\rangle =e^{i \vect k\cdot\vect R} |_{t, \vect k}^{n,j}\rangle$, form a plane wave basis. Since the non-abelian group $G_M$ has $q$-dimensional irreducible representations \cite{brown_physrev_1964}, the spectrum of such plane-wave states is $q$-fold degenerate.
Below, we use a basis of Bloch functions built from the LLs in Eq.~\eqref{eq:H_LL},  
\begin{align}
 |_{t, \vect k}^{n,j}\rangle =\!
  \frac{1}{\sqrt N}\!\sum_{s} e^{-ik_1 qas}\psi_n^{ k_2+\frac{\sqrt3}{2}b(ps+j+\frac{tp}{q})},\quad j=1,\cdots p \nonumber
\end{align}
where the sum runs over $s=-N/2, \cdots ,N/2$,  ($N\to\infty$).
This basis is similar to the set of Bloch states for a one-dimensional chain with $p$ sites per elementary unit cell, and multiple atomic orbitals on each site labeled by the LL index, $n$.
One more index $t\!=\!0,\cdots q\!-\!1$ takes into account the above-mentioned $q$-fold degeneracy, and  $\vect k\!=\!k_1 \hat{\vect k}_1+k_2 \hat{\vect k}_2$, $|k_i|\!<\!\frac{\sqrt3}{4q}b$, $ \hat{\vect k}_i\cdot\frac{\vect a_j}{a}=\delta_{i,j}$.

Then, we calculate the matrix elements,
\begin{align}
 &\langle_{t,\vect k}^{n,j}| H |_{\tilde t,\tilde{\vect k} }^{\tilde n ,\tilde j} \rangle     
 =\delta_{t,\tilde t} \delta_{\vect k ,\tilde{\vect k}} \sum_{s}  e^{-i k_1 q a s}    \mu _{n \tilde n }^{ k_2+\frac{\sqrt3}{2}bj, \, k_2+\frac{\sqrt3}{2}b(p s+\tilde j)}, \nonumber 
\end{align}
and find that the index $t$ is conserved, since $[\Theta_{\vect X},H]=0$ and 
 $\Theta_{\vect a_1}|_{t, \vect k}^{n,j}\rangle=|_{t+1, \vect k}^{n,j}\rangle$, 
 $\Theta_{\vect a_2}|_{t, \vect k}^{n,j}\rangle= e^{i\frac{2\pi p t}{q}} e^{i\vect k\cdot \vect a_2 } |_{t, \vect k}^{n,j}\rangle$.
Also, since Hamiltonian \eqref{eq:H} contains only the simplest moir\'e harmonics, the sum in the above expression is limited to $s\!=\!0,\pm 1$,
\begin{widetext}
\begin{align*}
&\mu _{n \tilde n }^{k_2\tilde k_{2}}
=\delta_{n,\tilde n}\delta_{k_2,\tilde k_2}E_n
+ \xi \beta v \sum _{ m } (u_1^+(-1)^m- i u_1^-)    
[e^{i 2\pi/3} s_{\tilde{n}}  (i b_m^1  +   b_m^2)  M^{n^-\!,\tilde{n}^+}_{m }\!   +       
e^{-i 2\pi/3} s_{n}            (- i b_m^1 +  ib_m^2)   M^{n^+\!,\tilde{n}^-}_{m }]\\
&\qquad\quad\; + vb \sum _{ m } (u_0^+ + i (-1)^m u_0^-)     
[ M^{n^-\!,\tilde{n}^-}_{m } \! +        s_{n} s_{\tilde{n}} M^{n^+\!, \tilde{n}^+}_{m}]
+ \xi vb   \sum _{ m } (u_3^- + i (-1)^m u_3^+)     
[ M^{n^-\!,\tilde{n}^-}_{ m}  \! -  s_{n} s_{\tilde{n}} M^{n^+\!, \tilde{n}^+}_{m}]\\
&M_{0 }^{ n_1\!,n_2} \!\!=\!   e^{i \beta \lambda^2 \!\sqrt{3}} \delta^- W_{1,\text{-}1,\text{-}1}^{n_1\!,n_2 },\; 
  M_{1 }^{n_1\!,n_2} \!\!=\!  \delta^- W_{\text{-}1,1,0 }^{n_1\!,n_2 } ,\;
  M_{2 }^{n_1\!,n_2} \!\!=\!  \delta^0W_{\text{-}1,0,1 }^{n_1\!,n_2 }   ,\;
  M_{3 }^{n_1\!,n_2} \!\!=\!  e^{i \beta \lambda^2  \!\sqrt{3}}\delta^+ W_{\text{-}1,\text{-}1,1}^{n_1\!,n_2 } ,\;
  M_{4 }^{n_1\!,n_2} \!\!=\!  \delta^+ W_{1,1,0}^{n_1\!,n_2 } ,\\
  &M_{5}^{n_1\!,n_2} \!\!=\!  \delta^0W_{1,0,\text{-}1}^{n_1\!,n_2 } , \quad
W_{c_1,c_2,c_3}^{n_1,n_2 }  =  \frac{N_{n_1,n_2}}{3}\lambda_B  A_{n_1}    A_{n_2}  2^{\bar{n}} 
 \sqrt{\pi}  \underline{n}! L_{\underline{n}}^{|\delta\!n|}\! (2\lambda ^2)\,
e^{\text{-}\lambda ^2}  \lambda^{|\delta\!n|} e^{i|\delta\!n|(c_1\!\frac{ \pi}{2} +c_2 s_{\!\delta\!n} \beta \!\frac{\pi}{3})} e^{i \beta c_3  \lambda^2 4 k_2/b   } 
\end{align*}
\end{widetext}
Here $L_n^{\alpha}\!(x)$ is the associated Laguerre polynomial, $\lambda \!=\!b \lambda_B/2$,  $\delta^{\pm}\!=\!\delta_{\tilde{k}_2,k_2\pm\sqrt{3}b/2}$, $\delta^{0}\!=\!\delta_{\tilde{k}_2,k_2}$,  $\bar{n}\!=\!\text{max}[n_1,n_2]$, $\underline{n}\!=\!\text{min}[n_1,n_2]$, $\delta\!n\!=\!n_1-n_2$, $s_n\!=\!\text{sign}(n)$, and $N_{n_1\!,n_2}\!=\!(1\!-\!\delta_{n_1\!,\text{-}1})\!(1\!-\!\delta_{n_2\!,\text{-}1})\!\sqrt{1\!+\!\delta_{n_1\!,0}}\sqrt{1\!+\!\delta_{n_2\!,0}}$.

Figure \ref{fig:miniband_spectra} shows a typical fractal spectrum of magnetic minibands, as a function of the magnetic flux $\phi=SB$ per moir\'e supercell: this one calculated for the inversion-symmetric superlattice perturbation with $V^+\!=\! 0.063vb$ and $ V^-\!=\!0$, $\theta\!=\!0$, by diagonalizing $H$, in the basis $|_{t, \vect k}^{n,j}\rangle$ (with a large enough range of $n$ to ensure convergence).
The $B\!=\!0$ miniband spectrum is displayed on the left, with the main DP and a secondary DP in the valence band.
For $\phi\lesssim0.2\phi_0$, the magnetic miniband spectra can be traced to the analytically calculated sequence of LLs associated with the two DPs.

At a higher flux, the LLs transform into a fractal spectrum with a hierarchy of bands and gaps, which can be understood upon analyzing the dispersion of electrons at the simplest flux fractions $\phi\!=\!\phi_0p/q$ (e.g.~$\phi\!=\! \smfrac{1}{2}\phi_0$, $\smfrac{4}{3}\phi_0$, $\smfrac{3}{2}\phi_0$)  and integer flux (e.g.~$\phi\!=\!2\phi_0$) shown as insets in Fig.~\ref{fig:miniband_spectra}. We find that the edges of many of the consecutive  bands can be described in terms of a weakly gapped Dirac spectrum, with $\Delta$ and $v_{\text{eff}}$ determined by a fit. 
Then, the largest gaps in the surrounding fractal spectrum at $B\!=\!B_{p/q}+\delta\!B$ can be interpreted \cite{chang_prl_1995} in terms of weakly broadened LLs of these next generation Dirac electrons, evaluated using Eq.~\eqref{eq:gapped_En} and parameters $\Delta$, $v\!\rightarrow \!v_{\text{eff}}$, and $B\!\rightarrow\!\delta\!B$. The analytically calculated effective LLs are shown in blue in the vicinity of the corresponding rational flux values. They interpolate the groups of minibands into the interval where the exact diagonalization becomes impractical, due to the basis set size. This correspondence can be seen best in the cases when the gap $\Delta$ between consecutive minibands is small, and the largest gaps are around the broadened ``$n\!=\!0$'' LL of the corresponding effective Dirac model.
The presence of Dirac-type features indicates that the corresponding 2D system in a magnetic field has the properties of a topological insulator, with chiral edge-states inside the band gaps responsible for the formation of the quantum Hall effect.
 
\begin{figure}[t]
\center
\includegraphics[width=0.5\textwidth]{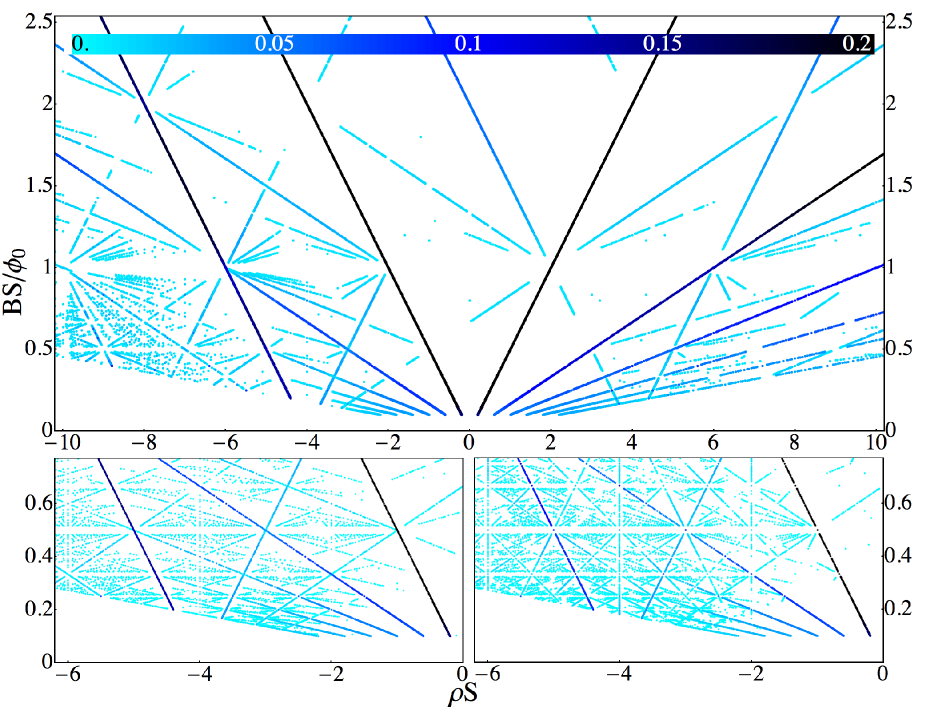}
\caption{
Density-magnetic field diagram showing incompressible electrons states in the fractal spectrum of magnetic minibands. Larger gaps are shown using a darker color. The two panels at the bottom are used to demonstrate the effect of an inversion asymmetric perturbation.
}
\label{fig:fan_diagram}
\end{figure}

In the upper and lower left panels of Figure \ref{fig:fan_diagram} we produce a map of incompressible states formed  upon filling the minibands in Fig.~\ref{fig:miniband_spectra}, taking into account that each magnetic miniband at $B_{p/q}$ accommodates a density $\delta\!\rho_{p/q}=g_sg_v/(qS)$  of electrons, and using darker color to reflect larger gaps in the fractal spectrum. As a reference, we use the largest gaps around the broadened ``$n\!=\!0$'' LL traced to the main DP in graphene. This is shown as the darkest lines corresponding to filling factors $\nu=\pm2$. Other dark lines correspond to the large gaps surrounding the ``$n\!=\!0$'' LLs in the next generation of Dirac electrons, e.g.~in the vicinity of $B_1$, $B_{1/2}$. They form fan-plots $\rho(B)$ in which the gradient changes at 
inter-crossings by
\begin{align*}
  \delta\!\!\left[\frac{d\!\rho}{d\!B}\right]=\frac{g_sg_v q}{\phi_0}
\end{align*}
reflecting the additional $q$-fold degeneracy of states in magnetic minibands at $B_{p/q}$ ($g_s\!=\! g_v\!=\!2$ account for spin-valley degeneracy). 
For example the darkest lines change the slope by $\delta[d\rho/dB]\!=\!8/\phi_0$ when crossing at $B_{1/2}$, in contrast to $\delta[d\rho/dB]\!=\!4/\phi_0$ at $B_1$.

Having investigated many different realizations of the hexagonal moir\'e superlattice in Eq.~\eqref{eq:H}, we conclude that robust Dirac-like features in the magnetic miniband spectrum have a systematic appearance.  
To mention, a Dirac-like band has been found in the Hofstadter spectrum of a tight-binding model on a square lattice with flux $\phi=\smfrac{1}{2}\phi_0$ per unit cell \cite{hou_pra_2009,delplace_prb_2010}. We also analyzed different realizations of hexagonal moir\'e superlattices described by Eq.~\ref{eq:H}, and found that the only qualitative difference comes from the inversion symmetry breaking in the unit cell:
the interplay between spatial and time inversion asymmetries lifts the valley degeneracy \cite{footnote:inv_sym} of the fractal spectra.
The lower-right panel of Fig.~\ref{fig:fan_diagram} offers an  example of the incompressibility map calculated for an inversion-asymmetric perturbation in Eqs.~(\ref{eq:H}-\ref{eq:model}) with $V^-\!=\!0.1V^+$. Here the Dirac-like features and the corresponding hierarchy of gaps in the magnetic minibands persist, but the fan diagram acquires additional lines, intercrossing with $\delta[d\!\rho/d\!B]\!=\!g_v q/\phi_0$.

{\it Acknowledgements}. We thank A.~Geim, P.~Jarillo-Herrero, P.~Kim, K.~Novoselov and D.~Weiss for useful discussions. This work was funded by the EU Flagship Project, EPSRC Science and Innovation Award and Grant EP/L013010/1, and ERC Synergy Grant Hetero2D.


\begin{thebibliography}{99}
\bibitem{footnote:inv_sym} 
For $u^-_i\!=\!0$, the spatial inversion symmetry, $H(\vect r,\zeta)\!=\!\sigma_zH(-\vect r,-\zeta)\sigma_z$, prescribes the relation $\epsilon_{\vect K+\vect k}\!=\!\epsilon_{\vect K'-\vect k}$ between spectra in graphene's two valleys. 
For $B\!=\!0$, time reversal symmetry prescribes the same relation, however when both $B\!\neq\!0$ and $u^-_i\!\neq \!0$ the spectra in the two valleys are unrelated.
\end{thebibliography}
\end{document}